\documentclass[11pt]{article}

\usepackage[margin=.75in]{geometry}
\usepackage{multirow}
\usepackage{datetime}
\usepackage[dvipsnames]{xcolor}
\usepackage{amsmath, amsthm, amssymb,fancyhdr,enumerate}
\usepackage{latexsym}
\usepackage[round, authoryear]{natbib}
\usepackage{hyperref}
\usepackage{MnSymbol}
\usepackage{graphicx}
\graphicspath{ {./images/} }
\title{Figures}

\renewcommand{\P}{\mathbb{P}}

\newcommand{\E}{\mathbb{E}}
\newcommand{\I}{\mathbb{I}}

\newcommand{\ah}{\textcolor{black}}

\begin{document}

\begin{center}
      \Large{\bf { Analysis of Ordinal Populations from Judgment Post-Stratification}}
\end{center}
\begin{center}
  \noindent{ {\sc Amirhossein Alvandi$^{\dagger}$}    and {\sc Armin Hatefi$^{\ddagger}$
  \footnote{Corresponding author: \\ Email: ahatefi@mun.ca, Tel: 709-864-8416}
            }}

\vspace{0.5cm}
\noindent{\footnotesize{ $^{\dagger}$ \it Department of Mathematics and Statistics, University of Massachusetts, Amherst, USA. \\
$^{\ddagger}$ \it Department of Mathematics and Statistics, Memorial University of Newfoundland, St. John's, NL, Canada. \\
}}
\end{center}

\begin{center} {\small \bf Abstract}: 
\end{center}

In surveys requiring cost efficiency, such as medical research, measuring the variable of interest 
(e.g., disease status) is expensive and/or time-consuming; However, we often have access to easily 
attainable characteristics about sampling units. These characteristics are not typically employed 
in data collection process. Judgment post-stratification (JPS) sampling enables us
to supplement the random samples from the population of interest with these characteristics as ranking information.
In this paper, we develop methods based on JPS samples for the estimation of categorical ordinal populations. 
We develop various estimators from JPS data even for a situation that JPS suffers from empty strata. 
 We also propose JPS estimators using multiple ranking resources. Through extensive numerical studies, we evaluate the performance of the methods in the estimation of the population.  Finally, the developed 
 estimation methods are applied to bone mineral data to estimate the bone disorder status of women aged 50 and older. 


\section{ Introduction } \label{sec:intro}
In survey sampling problems, we typically have various information (e.g., patients' demographic characteristics) 
about the population. Despite the accessibility, these characteristics are not often used in 
the data collection process. We believe these characteristics can be used for ranking and 
post-stratify the sampling units. The information, supplemented by the post-stratification, 
enables us to produce more representative samples from the population. 
     
   In bone disorder research, the bone mineral density (BMD) of patients is measured through 
   dual X-ray absorptiometry (DXA) imaging, a costly and time-consuming procedure. Although 
   there are plenty of patients, BMD measurements are obtained from a small proportion of 
   the population. In these situations, we randomly select sampling units from the population 
    and form a comparison set for each examined patient. As BMD measurements are expensive, we use 
    easily attainable characteristics (such as age, weight or BMD measurements from previous examinations)
    to rank the units (without measuring their BMD values) in their comparison sets.
   Through this sampling scheme, we can supplement the 
   BMD observations with these extra judgment ranks. This extra information 
   (obtained from rank-based stratification) enables us to better estimate patients' bone disorder status. 
       
     Judgment post-stratification (JPS) sampling \citep{maceachern2004judgement} and ranked set sampling
      (RSS) \citep{mcintyre1952method}) are two cost-effective rank-based sampling designs. In RSS, the 
      ranks of the measured units have to be pre-specified. Moreover, RSS data are collected after the 
      ranking process. Unlike RSS, JPS enjoys post-stratification so that observations are obtained before ranking. 
 JPS scheme offers various advantages over its RSS counterpart. For example, if the ranking process is unreliable, 
 one can simply ignore the attached ranks to the JPS  samples.  The JPS samples now form simple random samples (SRS) 
 from the population. Thus, one can apply the conventional statistical methods for analyzing the JPS data. Unlike 
 JPS, RSS statistics are not identically distributed. Hence, ranks can not be separated from the RSS observations \citep{chen2013ranked}.

           The following
 explains how one can obtain the judgment ranks and JPS data from the bone disorder example.
           Let $X_1,\ldots, X_n$ be an initial SRS of size $n$ representing 
the bone disorder status of patients $u^*_1,\ldots,u_n^*$. We shall supplement this 
initial SRS with judgment ranks and construct JPS data of size $n$. 
As measuring the bone status is costly, we only measured the bone status (i.e., $x$-value) 
of these $n$ patients as our initial SRS. To assign a judgment rank to the
$u^*_i$, we first take a sample of  $H-1$  patients   $u_{i,1},\ldots, u_{i,H-1}$ 
independently from the population and construct the comparison set for $u^*_i$, say 
${\bf S}_i=\{u^*_i,u_{i,1},\ldots, u_{i,H-1}\}$. Note that we do not measure the bone 
status of these $H-1$ patients. 
We then rank the patients in the ${\bf S}_i$ based on an easy-to-measure 
characteristic such as the age of patients.  
Let $R_i$ denote the judgment rank of  $u^*_i$ in the ${\bf S}_i$. Consequently, we 
consider $(X_i,R_i)$ as a JPS observation for patient $u^*_i$ where $X_i$ is 
the bone disorder status of the patient and $R_i$ is the judgment rank assigned to $X_i$ 
using the age characteristic. Finally, we replicate the above procedure to obtain the 
JPS data $\{(X_i,R_i),i=1,\ldots,n\}$ for the bone disorder population.

In the JPS scheme,  we supplement the initial SRS $(X_1,\ldots, X_n)$ with judgemental ranks $(R_1,\ldots,R_n)$ using easy-to-measure characteristics.  
Because the judgment rank $R_i$ may differ from the true rank of $X_i$ in the $i$-th comparison set, the ranking is called imperfect in this case. Unlike imperfect JPS, when units are ranked based on the response variable $X$, the $R_i$ is the true rank of $X_i$. In this case, there is no ranking error and the ranking is called perfect. 
             
      JPS sampling has been employed in many research studies and applications. 
      \cite{wang2006concomitants} used JPS samples for the estimation of the population 
      mean \cite{frey2012improved} investigated the admissibility of the JPS estimator 
      of the population mean. The isotonic regression model was used by \citep{wang2008nonparametric} 
      to develop a class of isotonized estimators for the cumulative distribution function. 
   \cite{ozturk2012combining} proposed a new class of estimators based on multi-ranker JPS. 
   \cite{ozturk2014statistical} utilized the properties of JPS samples for non-parametric inference of population quantiles. 
   \cite{dastbaravarde2016some} focused on parametric inference from JPS samples. \cite{omidvar2018judgment} 
   estimated finite mixture models with JPS samples
   \cite{zamanzade2017estimation} studied the estimation problem of a binary population using JPS data. 
   In the literature of the categorical ordinal variables, various research studied the estimation of the
    ordinal population from RSS samples. \cite{chen2008ranked} utilized the ordinal logistic regression 
    for the estimation of the ordinal population with RSS data. \cite{hatefi2020efficient,alvandi2021estimation} used RSS data 
    with tie-structures to estimate the ordinal population parameters. Despite the importance of categorical 
    ordinal variables and the challenges of the RSS \citep{wang2008nonparametric}, to our best knowledge, 
    no research in the literature explored the estimation problem of ordinal populations with JPS data. 
    In this manuscript, we investigate the properties of JPS data for the estimation of categorical ordinal populations. 
         
       This manuscript is organized as follows. In Section \ref{sec:est}, we propose various estimation methods 
       for ordinal population parameters. These methods include three estimators from JPS without empty strata 
       and six JPS estimators with empty strata. 
 In Section \ref{sec:multi}, the estimation problem is extended to JPS data from multiple rankers.  
 In Section \ref{sec:sim}, through extensive simulation studies, we evaluate the performance of the estimators.
   The proposed methods are applied for the analysis of the bone disorder population of older people in Section 
   \ref{sec:real}. Finally, we present the summary and concluding remarks in Section \ref{sec:sum}. 
        
\section{Estimation Procedures}\label{sec:est}

Let  $X$ follow a multinomial distribution with $Q$ ordinal categories, i.e., $X \sim \text{Multi}(1,p_1,\ldots,p_Q)$ with $\sum_{q=1}^{Q} p_q=1$  and $p_q >0, q=1,\ldots,Q$. Also let $c_q, q=1,\ldots,Q$ denote the cumulative probabilities where $ c_q= P(X \le q) = \sum_{l=1}^{q} p_l$ with $c_0 \equiv 0$ and $c_Q = 1$. 
Note that $X$ represents one draw out of $Q$ ordinal categories of the multinomial distribution. Thus $X$ is a univariate random variable taking values on $\{1,2,\ldots,Q\}$.
Let ${\bf X}_{jps}=\{X_1,\ldots,X_n\}$ represent a JPS sample of size $n$ from $X$-population with judgment ranks ${\bf R}=(R_1,\ldots,R_n)$ 
where $R_i$ follows discrete uniform distribution on set $\{1,\ldots,H\}$.
In this section, we propose various estimation methods from JPS data for population proportions.  

\subsection{Standard Non-parametric Estimators} \label{sub:srs}
We  consider the estimation of ordinal population based on commonly-used  simple random sample/sampling (SRS).
Note when we ignore the ranks of JPS data, ${\bf X}_{jps}$ can be treated as an SRS of size $n$. 

Let $Y_q, q=1,\ldots,Q$ denote the number SRS data obtained from categories $\{1,\ldots,q\}$; that is
 $Y_q = \sum_{i=1}^{n} \I(X_i\le q)$,
 where $\I(\cdot)$ represents an indicator function that $\I (X_i \le q)=1$ if $X_i \le q$, otherwise 
 $\I (X_i \le q)=0$.  
 Suppose ${\bf \mathcal Y}_{srs}=(Y_1,\ldots,Y_Q)$ and ${\bf c}=(c_1,\ldots,c_Q)$.
 The standard  estimator of $c_q$
from the SRS is given by
\begin{align}
{\widehat c}_{q,srs} = \frac{Y_q}{n}, ~~~~ \forall q=1,\ldots, Q-1.
\end{align}
 The population proportions  are finally estimated by ${\widehat p}_{q,srs} = {\widehat c}_{q,srs} -{\widehat c}_{q-1,srs}$ for $q=1,\ldots,Q$. 
We now focus on estimation of the parameters from JPS data. 
Let $n_h=\sum_{i=1}^{n} \I(R_i=h)$ represent the number of JPS data obtained from rank stratum $h$.
Let $p_{[h]q}$ be the probability that a JPS observation with rank $h$ comes from category $q$. 
Hence, 
\[
p_{[h]q}=\P(X_i=q | R_i=h),  ~~~ q=1,\ldots, Q,
\]
where $\sum_{q=1}^{Q} p_{[h]q}=1$ for any $h, h=1,\ldots, H$.  Similarly, let  
$c_{[h]q}$ be the probability that a JPS observation with rank $h$ comes from categories $\{1,\ldots,q\}$. 
Hence, 
\[
c_{[h]q}=\P(X_i \le q | R_i=h),  ~~~ q=1,\ldots, Q.
\]
Let ${Y}_{[h]q}$ denote the number of JPS data from $h$-th rank stratum from categories $\{1,\ldots,q\}$; 
that is
${Y}_{[h]q}= \sum_{i=1}^{n} \I(X_i \le q|R_i=h),\forall q=1,\ldots, Q$. 
We define ${\bf \mathcal{Y}}_{[h]}=\left( {Y}_{[h]1}, {Y}_{[h]2}, 
\ldots, {Y}_{[h]Q}\right)$ and
${\bf c}_{[h]}=(c_{[h]1},c_{[h]2},\ldots,c_{[h]Q})$.
Hence, the standard JPS estimator of $c_{[h]q}$ is given by
\begin{eqnarray}\label{prq_st}
{\widehat c}_{[h]q,st}=\frac{{Y}_{[h]q}}{n_h} 
 = \frac{1}{n_h}\sum_{i=1}^{n}\I(X_i\le q|R_i=h).
\end{eqnarray}
Using \eqref{prq_st}, the standard JPS estimator of $c_{q}$ is obtained by
\begin{eqnarray}\label{pq_st}
{\widehat c}_{q,st}= 
\frac{1}{\sum_{h=1}^{H} \I(n_h > 0)}  
\sum_{h=1}^{H} {\widehat c}_{[h]q,st} \I(n_h > 0), ~~ q=1,\ldots,Q-1.
\end{eqnarray}
 The population proportions under standard JPS are finally estimated by ${\widehat p}_{q,st} = {\widehat c}_{q,st} -{\widehat c}_{q-1,st}$ for $q=1,\ldots,Q$. 

 \subsection{Maximum Likelihood Estimator}\label{sub:ml_rsst}
We use the maximum likelihood (ML) method to estimate 
the ordinal population proportions based on JPS data.
Let  $X_1,\ldots,X_H$ be an SRS of size $H$ from an ordinal distribution with categories $\{1,\ldots,Q\}$ where $X \sim 
 \text{Multi}(1,p_1,\ldots,p_Q)$. Let $X_{(h)}$ represent the $h$th order statistic
   in a set of size $H$ units. For $h=1,\ldots, H$, following \cite{david2004order}, 
   one can easily show that $X_{(h)} \sim \text{Multi}(1,p_{(h)1},\ldots,p_{(h)Q})$ with
\begin{align}\label{prq} \nonumber 
p_{(h)q} &= \frac{1}{B(h,H-h+1)} \int_{c_{q-1}}^{c_{q}} u^{h-1} (1-u)^{H-h} du\\
&= B_{c_q}(h,H-h+1) - B_{c_{q-1}}(h,H-h+1),
\end{align}
where 
\[
B_{c_q}(h,H-h+1) = \frac{1}{B(h,H-h+1)} \int_{0}^{c_{q}} u^{h-1} (1-u)^{H-h} du,
\]
denotes the  incomplete beta function for $q=1,\ldots,Q$ with  $c_0 \equiv 0, c_Q = 1$.

Let $ {\bf X}_{jps}=(X_1,\ldots,X_n)$ with ranks ${\bf R}=(R_1,\ldots,R_n)$ denote a perfect JPS sample (i.e., ranking is done based on $X$ values and there is no ranking error) of size $n$ from 
the population.
Given $R_i=r_i$, $X_i$ represents the $r_i$-th order statistic in the set of $H$ units. 
Thus the conditional log-likelihood function of ${\bf X}_{jps}$ can be constructed as follows (where the proof can be found in the Appendix):
\begin{align}\label{ll}
{\ell}({\bf c})=l(c_1,\ldots,c_{{Q-1}})=\sum_{i=1}^{n} 
\left\{ 
 \I(X_i=1) \log(p_{(r_i)1}) + \ldots +\I(X_i=Q) \log\left(1- \sum_{q=1}^{Q-1}p_{(r_i)q}\right)  
\right\},
\end{align}
 where $p_{(r)q}, \forall r, q$ as incomplete beta function of $(c_1,\ldots,c_{Q-1})$ is given by \eqref{prq}. Hence, the log-likelihood function is written as a function of $c_1,\ldots,c_{Q-1}$ on the left-hand side of  \eqref{ll}.
 From \eqref{ll},
 the ML estimator of
$c_q$ from JPS data is given by 
\begin{align}\label{pq_ml}
{\bf \widehat c}_{ml}= ({\widehat c}_{1,ml},\ldots, {\widehat c}_{Q-1,ml}) = 
\underset{ 0 < c_1 < \ldots < c_{Q-1} < 1
 }{{\arg}\max} l({\bf c}),
\end{align}
the constraint $0 < c_1 < \ldots < c_{Q-1} < 1$ is implied by the fact that $Q$ is given and $p_q>0,q=1,\ldots,Q$ in the population. To meet this requirement, throughout the manuscript  for all the estimation methods including the ML method, we only focus on the samples with at least one observation from all categories in the initial SRS. 
To find the maximum of log-likelihood function \eqref{ll}, we  treated the $p_{(r)q}$ as incomplete beta functions of $(c_1,\ldots,c_{Q-1})$. We then  used the linearly constraint optimization  via constrOptim command in R subject to constraints  $0 < c_1 < \ldots < c_{Q-1} < 1$ and $0.01 \le c_q \le 0.99$ for $q=1,\ldots,Q-1$
 to estimate a legitimate multinomial distribution in the numerical optimizations. We used the estimates from the standard JPS method as the starting values. The stopping rule was built by the convergence tolerance $10^{-9}$ and maximum number of iterations $1000$.
Note that we developed the ML estimators assuming  that there is no
 ranking error in the JPS; however, ranking error is undeniable  
 in real life applications where ranking is carried out 
 through auxiliary variables.
 Accordingly, we assess  the performance of the estimators 
   in presence of ranking errors in both simulation and real data analysis. 

\subsection{Non-parametric Estimation Procedures }\label{sub:non}
We focus on isotonized version of the standard estimator and develop new non-parametric estimation procedures for ordinal population using JPS data in presence and absence of empty strata.

\subsubsection {\bf JPS  Data without Empty Strata} \label{sub:iso_no}
We first study the isotonized estimation of the population proportions assuming that no empty stratum is observed in JPS data. 
Since it is more likely that the statistics with smaller ranks come from smaller categories, it is reasonable to impose the constraint 
\begin{align}\label{order}
c_{[1]q} \ge c_{[2]q} \ge \ldots c_{[H]q}, ~~~ q=1,\ldots,Q.
\end{align}
 The constraint \eqref{order} may be violated owing to the sampling variability in JPS data.
 One can use the isotonized strategy proposed by \cite{ozturk2007statistical} to impose the constraint \eqref{order} in aim to minimize the weighted least square. \cite{wang2008nonparametric} and \cite{zamanzade2017estimation} utilized the strategy in the estimation of the population mean and the binary proportion, respectively. We apply the isotonized strategy to estimate the ordinal population with JPS data.

When the JPS has no empty strata, the isotonized estimator of $\{c_{[r]q}, r=1,\ldots,H\}$ can be obtained by  either the MinMax method
\begin{align}\label{niso-}
{\widehat c}_{[r]q,iso^-} = 
 \underset{ h \le r }{\min} ~ \underset{s \ge r}{\max} \sum_{g=h}^{s} 
\frac{n_g ~ {\widehat c}_{[g]q,st}}{n_{hs}},
\end{align}
or the MaxMin isotonized estimator 
\begin{align}\label{niso+}
{\hat c}_{[r]q,iso^+} = 
\underset{s \ge r}{\max} ~ \underset{ h \le r }{\min} ~  \sum_{g=h}^{s} 
\frac{n_g ~ {\widehat c}_{[g]q,st}}{n_{hs}},
\end{align}
where $n_{hs}=\sum_{g=h}^{s} n_g$ and $n_g=\sum_{i=1}^{n} \I(R_i=g)$.
Therefore, for JPS data without empty strata, an isotonozed estimator of the cumulative probability $c_q$ is given by
\begin{eqnarray}\label{p_niso}
{\widehat c}_{q,iso}=\frac{1}{H} \sum_{r=1}^{H} {\widehat c}_{[r]q,iso}, ~~~ q=1,\ldots,Q-1.
\end{eqnarray}
 When there is no empty strata, ${\widehat c}_{[r]q,iso^-}$ and ${{\widehat c}_{[r]q,iso^+}}$ yield an 
identical result. Hence, one can obtain \eqref{p_niso} using either ${{\widehat c}_{[r]q,iso^-}}$ or ${{\widehat c}_{[r]q,iso^+}}$ 
as ${{\widehat c}_{[r]q,iso}}$. To compute ${{\widehat c}_{q,iso}}$, one can first use isotonized strategy for
 ${\bar Y}_{[r],q},~q=1,\ldots,Q$ via the Pool Adjacent Violator Algorithm (PAVA) and then take the average. {From  ${\widehat c}_{q,iso}$, one can find the isotonized estimator of the population proportion $p_q$ by  ${\widehat p}_{q,isp} = {\widehat c}_{q,iso} -{\widehat c}_{q-1,iso}$ for $q=1,\ldots,Q$}.

\subsubsection {\bf JPS  Data with Empty Strata}
Now, we investigate the estimation of ordinal population from JPS with empty strata.
 We propose four isotonized estimation procedures including naive estimator ignoring the empty strata, MinMax, MaxMin and mixed estimators. 
When JPS contains empty strata,  \eqref{niso-} and \eqref{niso+} may not be the same. 
In the presence of empty strata, the pooled cells (i.e., $n_{rs}$) used in \eqref{niso-} and \eqref{niso+} may still be empty. 
\cite{wang2008nonparametric} proposed the idea of index set 
$l(u)=\{v| \sum_{k=\min(u,v)}^{\max(u,v)} n_k > 0, 1 \le v \le H \}$ to deal with the empty strata. 
The MinMax and MaxMin estimators in presence of empty strata  are given by
\begin{align}\label{yiso-}
{\widehat c}_{[r]q,iso^-} = 
 \underset{ h \le r, r\in l(H) }{\min} ~ \underset{s \ge r, s\in l(r)}{\max} \sum_{g=h}^{s} 
\frac{n_g ~ {\widehat c}_{[g]q,st}}{n_{hs}},
\end{align}
\begin{align}\label{yiso+}
{\widehat c}_{[r]q,iso^+} = 
~ \underset{s \ge r, s\in l(1)}{\max} ~ \underset{ h \le r, r\in l(s) }{\min}  \sum_{g=h}^{s} 
\frac{n_g ~ {\widehat c}_{[g]q,st}}{n_{hs}},
\end{align}
where $n_{hs}=\sum_{g=h}^{s} n_{g}$ and {${\widehat c}_{[g]q,st}$} are given by \eqref{prq_st}.

When JPS data has empty strata, the first approach can be simply an estimator ignoring the empty strata.  
After dropping the empty strata, one can obtain the isotonized estimate of  the proportions in a similar 
fashion as Subsection \ref{sub:iso_no}. This estimator is also denoted by ${\widehat c}_{q,iso}$ to 
emphasize  that empty strata has been ignored in the isotonized strategy. 

Using \eqref{yiso-} and \eqref{yiso+}, we propose our next estimators for proportions from JPS with empty 
strata. The MinMax isotonized estimator is given by 
${\widehat c}_{q,iso^-} = \frac{1}{H} \sum_{r=1}^{H } {\widehat c}_{[r]q,iso^-}$ where ${\widehat c}_{[r]q,iso^-}$ are 
estimated from \eqref{yiso-}. In a similar vein,  we  use \eqref{yiso+} and propose the MaxMin estimator by 
${\widehat c}_{q}^{iso^+} = \frac{1}{H} \sum_{r=1}^{H } {\widehat c}_{[r]q,iso^+}$.
The empty strata in JPS may result in empty pooled strata (i.e., $n_{rs}$) in the isotonized strategy.
 Hence, the MaxMin and MinMax estimators differ when an empty stratum occurs. 
To compute the MinMax and MaxMin isotonized estimators, one can first implement PAVA algorithm over non-empty 
cells and obtains the corresponding in-stratum proportion estimates. Note that an empty stratum is located at the boundary,
 if the empty stratum is surrounded by empty strata on both sides.
If the empty stratum is not at the boundary, the MinMax imputes the value by the closest non-empty in-stratum estimate on
 its left while the MaxMin uses the closest non-empty in-stratum estimate on the right for imputation. 
When we face with empty cell at the boundary, then both MinMax and MaxMin estimators use  the closest available 
non-empty stratum to estimate the empty values. For more information, see \cite{wang2012isotonized}.
  
 \cite{zamanzade2017estimation} proposed an isotonized estimator for binary proportion, by simply averaging the MaxMin and MinMax estimators. Following this study, we finally combine the two estimators and propose a new estimator for ordinal population from JPS with empty strata. This combined estimator is given by 
 ${\widehat c}_{q,iso^*}=\frac{{\widehat c}_{q,iso^-} + {\widehat c}_{q,iso^+}}{2}$ where ${\widehat c}_{q,iso^-}$ and ${\widehat c}_{q,iso^+}$ are the MaxMin and the MinMax estimators. Finally, similar to Subsection \ref{sub:iso_no}, one can obtain the isotonized estimators of $p_q,q=1,\ldots,Q$ from their corresponding isotonized estimators for the cumulative probabilities.

\section{Extension to Multiple Rankers }\label{sec:multi}
In many applications, the ordinal variable is accompanied by several easy to measure concomitant variables. In these cases, multiple ranking information are available for each sampling unit. Therefore, it is important how to combine ranking information from multiple resources and use them in the estimation process.
In this section, we develop three estimators from JPS  with multiple rankers for ordinal population. These estimators include standard estimator, its isotonized counterpart and an logistic regression based estimator. 

\subsection{Standard  and Isotonized Multi-ranker Estimator}\label{sub:stan_multi}
Suppose that we have access to ranking information from $K$ rankers. 
Let ${\bf X}_{mp}=\left(X_i,{R}_{i1},\ldots,{R}_{iK}\right)_{i=1}^n$ represent JPS data of size $n$ 
where ${R}_{ik}$ is the judgment rank of $X_i$ assigned by ranker $k$ for $k=1,\ldots,K$.
Following \cite{wang2012isotonized},  we propose a standard estimation method based on multi-ranker JPS. Here, the ranking information is first averaged for each JPS observation and then used in estimation process. The standard estimator for {$c_q$  for $q=1,\ldots,Q-1$} is given by 
\begin{align}\label{pq_sm}
{{\widehat c}_{q,sm}}= \frac{1}{H} \sum_{h=1}^{H} {{\widehat c}_{[h],q,sm}}= 
\frac{1}{H} \sum_{h=1}^{H} \frac{\sum_{i=1}^{N} \I({X_i \le q}) \gamma_{ih}}{\sum_{i=1}^{N}\gamma_{ih}},
\end{align}

where 
$ \gamma_{ih} = \sum_{k=1}^{K} \delta_k \I({R_{ik}=h})$ with $\delta_k=|\rho_k|/\sum_{k=1}^{K} |\rho_k|$
where the ranking ability  $\rho_k=cor(X,Z_k)$ denotes the sample correlation between $X$ and ranker $k$, namely $Z_k$, for $k=1,\ldots,K$.

Although there are $K$ ranks available for each JPS observation here,  the constraint  \eqref{order} may still be violated by the standard multi-ranker estimator.  
Following \cite{wang2008nonparametric}, one can obtain the isotonized estimator of the population proportions from JPS with multiple rankers. Thus, the multi-ranker estimator of $c_q$ for $q=1,\ldots,Q-1$ with isotonic properties is given by 
\begin{align}\label{pq_ism}
{{\widehat c}_{q,sm*}}= \frac{1}{H} \sum_{h=1}^{H} {\widehat c}_{[h]q,sm*}
\end{align}
where ${\widehat c}_{[h]q,sm*}$ is the isotonized version of ${\widehat c}_{[h]q,sm}$ with weights $\{{\widetilde n}_h\}_{h=1}^{H}$ where
\[
{\widehat c}_{[r]q,sm*} = 
\underset{r \le h}{\max} ~ \underset{ s \ge r }{\min} ~  \sum_{g=r}^{s} 
\frac{{\widetilde n}_g ~ {\widehat c}_{[g]q,sm}}{{\widetilde n}_{hs}},
\]
where ${\widetilde n}_g = K \sum_{i=1}^{n} \gamma_{ig}$ and ${\widetilde n}_{hs} = \sum_{g=r}^{s} {\widetilde n}_g$.

\subsection{OLR Based Estimation}\label{sub:olr}
Another methodology is to benefit from the  ordinal logistic regression (OLR) model to estimate the proportions from  JPS with multiple rankers.  The OLR model is utilized to combine information from multiple rankers into the estimation process. 
Let ${\bf X}=(X_1^*,\ldots,X_n^*)$ be the initial SRS of size $n$ from which we shall obtain JPS data of size $n$. Also let ${\bf Z} = (Z_1,\ldots,Z_K)$ represent the vector of ranking variables corresponding to the initial SRS.
Following \cite{chen2008ranked}, 
we fit an OLR model with common slopes on the cumulative probabilities $c_q=\P(X \le q)$. The OLR model is given by
\[
\text{logit}(c_q) = \alpha_q + {\bf\beta}^\top {\bf Z}, ~~~  1 \le q \le Q-1,
\]
where $\text{logit}(c_q)=\log(c_q/1-c_q)$ with  slope ${\bf\beta}$  and
 $\alpha_1,\ldots, \alpha_{Q-1}$ the intercepts of the model.
In this approach, we first estimate the parameters of the OLR model using the initial SRS of size $n$. The trained OLR model is then used to carry out the ranking involved in the JPS sampling. 

We form a set of $H$ units ${\bf S}=\{u_1^*, u_1,\ldots,u_{H-1}\}$ including the first unit of the JPS data (say $u_1^*$ with variable of interest $X_1^*$) and $H-1$ randomly selected units $\{u_1,\ldots,u_{H-1}\}$ from the population (without measuring their variable of interest). 
 Applying the trained OLR model to the ${\bf S}$, we rank the units based on 
${\widehat \alpha}_q + {\widehat{\bf\beta}}^\top {\bf Z}$ where smaller values getting bigger 
ranks. This is because the bigger the value of ${\widehat \alpha}_q + {\widehat{\bf\beta}}^\top {\bf Z}$,
 the more likely it is that the unit should come from a smaller category.
Through this approach, we find the rank (say $R_1^*$) of  $X_1^*$. Similarly, we compute the ranks  of all JPS observations  using the trained OLR model. 
 Let ${\bf \tilde X}_{jps}=(X_1^*,\ldots,X_n^*)$ with ranks ${\bf \tilde R}_{jps}=(R_1^*,\ldots,X_n^*)$ denote the JPS from the OLR model. Thus the OLR-based JPS estimator of $c_q$ is given by
\begin{eqnarray}\label{pq_reg}
{{\widehat c}_{q,reg}}= 
\frac{1}{\sum_{h=1}^{H} \I(n_h > 0)}  
\sum_{h=1}^{H} {{\widehat c}_{[h]q,reg}} \I(n_h > 0), ~~ q=1,\ldots,{Q-1},
\end{eqnarray}
where
\begin{eqnarray*}\label{prq_reg}
{{\widehat c}_{[h]q,reg}} = \frac{1}{n_h}\sum_{i=1}^{n}\I({X_i^* \le q}|R_i^*=h).
\end{eqnarray*}
{Finally, similar to Subsection \ref{sub:iso_no}, the multi-ranker JPS estimators of $p_q,q=1,\ldots,Q$ can be derived from their corresponding  estimators of the cumulative probabilities}. 
\section{Simulation Studies}\label{sec:sim}
We investigate  the performance of the proposed methods in estimation of ordinal population. Let the variable of interest follow an ordinal population with three categories, i.e., 
$X \sim \text{Multi}(1,{\bf p})$ where ${\bf p}={(p_1,p_2,p_3)}$.
Here, we generate the ranking variable $Z$ such that  $X$ and $Z$  have correlation $\rho=cor(X,Z)$.
Given $\rho$, we generate $Z$ by
$ Z= \rho \left( \frac{X-\E[X]}{\text{SD}(X)}\right) + \epsilon  \sqrt{1-\rho^2} $ where $\E[X]$ is the mean for $X$, $\text{SD}(X)$ is the standard
deviation for $X$ and $\epsilon$ is a standard normal random variable that is independent of $X$.

In the simulation studies, we set  the correlation $\rho =\{0.5,0.7,0.9\}$ to better explore the effect of ranking errors on the estimators. To evaluate the effect of population configuration, we also set ${p_3}\in\{0,1,0.4,0.6\}$, ${p_2} \in \{0.1,\ldots,0.8\}$ and {$p_1$} was treated as the reference category. For each combination $(\rho,{\bf p})$, we generated JPS data size $n=\{30,60\}$ with set size $H=\{3,6\}$. For multiple rankers estimation, we generated two concomitant variables  $Z_1$ and $Z_2$ independently given $X$ such that $\rho_1=cor(X,Z_1)$ and $\rho_2=cor(X,Z_2)$. We considered three combinations of concomitant variables 
to generate JPS with multiple rankers.
These combinations include two strong rankers $(\rho_1,\rho_2)=(0.9,0.9)$, one strong ranker and one moderate ranker $(\rho_1,\rho_2)=(0.9,0.7)$ and two moderate rankers $(\rho_1,\rho_2)=(0.7,0.7)$. 
To compare the performance of the estimators, we used the total efficiency of the JPS estimators relative to their SRS counterparts of the same size. The total relative efficiency (RE) of a JPS estimator ${\widehat{\bf p}}_{jps}$ is given by 
\begin{align}\label{tre}
\text{RE}({\widehat{\bf p}}_{jps},{\widehat{\bf p}}_{srs})=\frac{\sum_{q=1}^{Q}\text{MSE}({\widehat p}_{q,srs})}{\sum_{q=1}^{Q}\text{MSE}({\widehat p}_{q,jps})},
\end{align}
where the RE is greater than one, we can conclude that the JPS estimator outperforms its SRS counterpart in the estimaipon of ordinal population.  
We carried out two separate simulations to investigate the performance of the estimators using JPS without empty strata and JPS with empty strata. In the first simulation, we only focused on the JPS data without empty 
strata. To do that, when the JPS sample was generated, we first checked if there was at least one 
observation from all rank strata. If not,  we ignored the sample and 
kept independently sampling  until we obtained JPS sample without empty strata. We then applied the developed methods to the JPS sample to estimate the parameters based on the JPS data without empty strata.
For JPS with empty strata, we carried out another simulation study. In this study, we generated a JPS sample and checked if there was at least one empty rank stratum. To do that, we first randomly selected the index of the empty strata. The proposed ranking design is accepted with probability 1/(number of empty strata) to favour sample with smaller number of empty strata. 
 We then kept independently sampling from the population until we obtained a JPS sample compatible with the accepted ranking design. The obtained JPS sample has at least one empty stratum. We then used the JPS sample with at least one empty stratum to estimate the parameters of the population.
For a fair evaluation,  we simulated  5000 times the JPS and SRS data and computed the estimates of the population parameters for each combination $(\rho,{\bf p},n,H)$.

\begin{figure}[h]
\begin{center}
\includegraphics[width=6in]{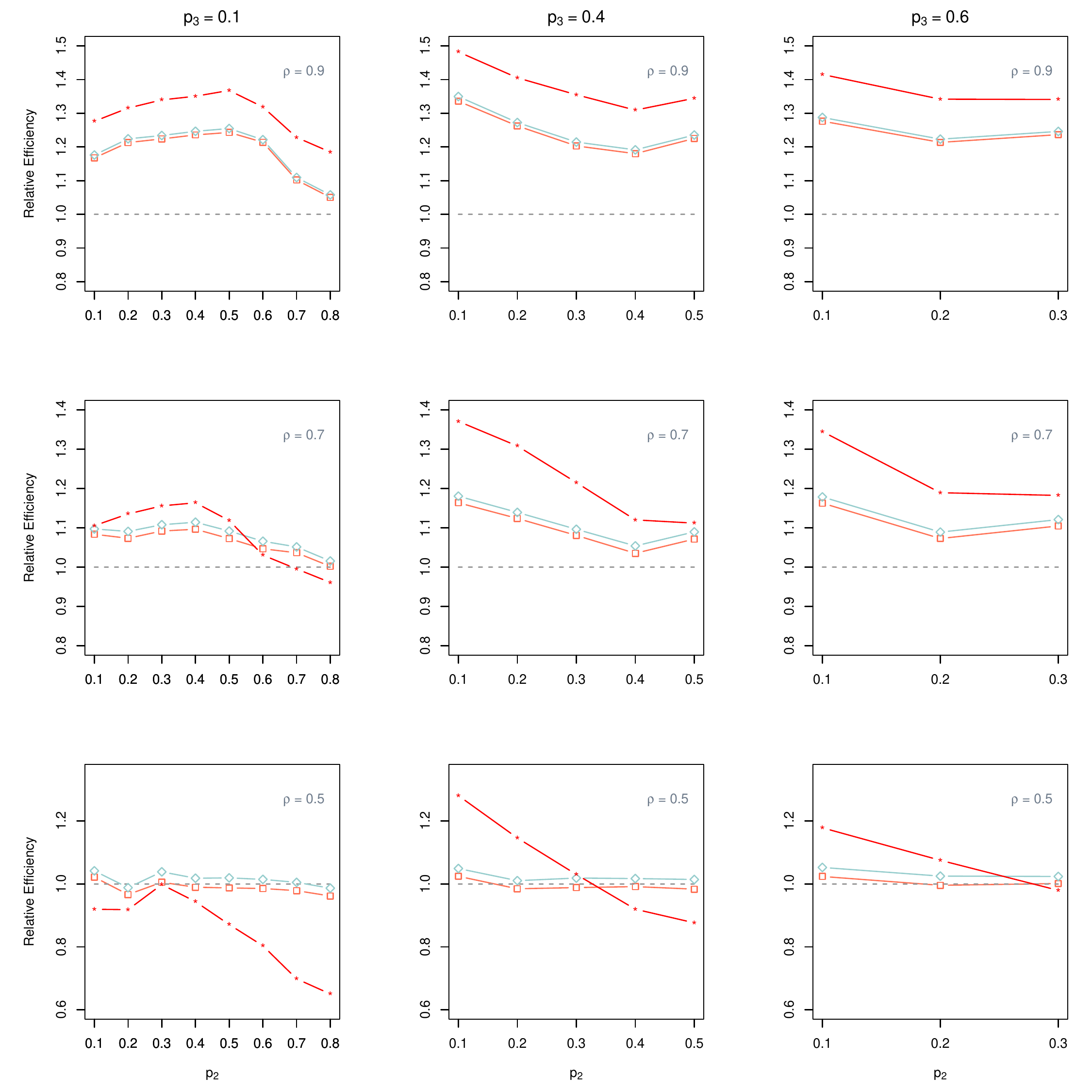}
\caption{\footnotesize{The total REs of $\widehat{\bf p}_{st}$ ({$\color{Melon}\medsquare$}), $\widehat{\bf p}_{iso}$ ({$\color{Aquamarine}\meddiamond$}), $\widehat{\bf p}_{ml}$ ({$\color{Red}\filledstar$}) from JPS without empty strata to their SRS counterpart  when $(n,H)=(30,3)$}.}
 \label{no_empty_N30m3}
\end{center}
\end{figure}

\begin{figure}[!ht]
\begin{center}
\includegraphics[width=6in]{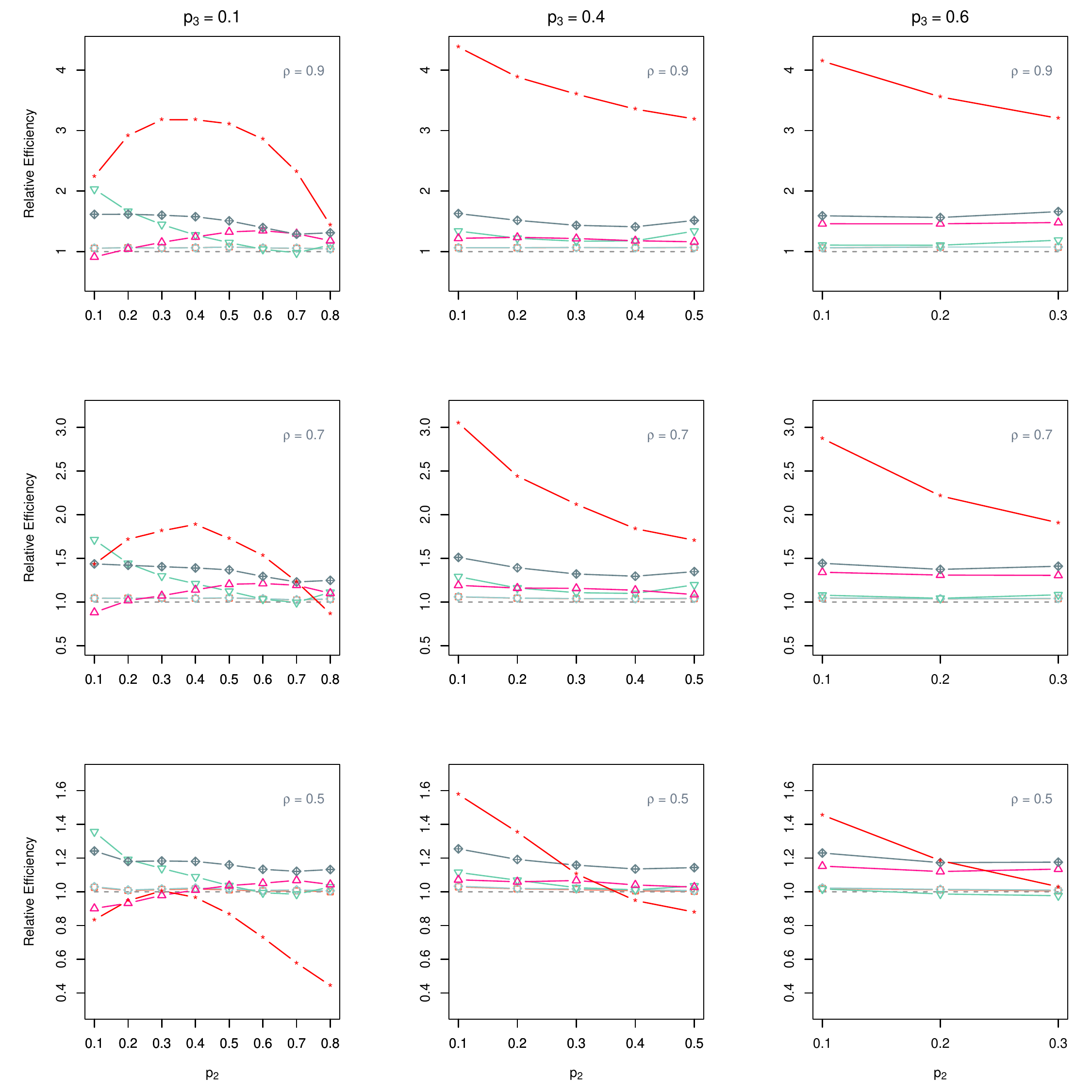}
\caption{\footnotesize{The total REs of $\widehat{\bf p}_{st}$ ({$\color{Melon}\medsquare$}), $\widehat{\bf p}_{iso}$ ({$\color{Aquamarine}\meddiamond$}), $\widehat{\bf p}_{q,{iso}^-}$ ({$\color{JungleGreen}\medtriangledown$}), $\widehat{\bf p}_{q,{iso}^+}$ ({$\color{RubineRed}\medtriangleup$}), $\widehat{\bf p}_{q,{iso}^*}$ ({$\color{RoyalBlue}\diamondplus$}),  $\widehat{\bf p}_{ml}$ ({$\color{Red}\filledstar$}) from JPS with empty strata to their SRS counterpart  when $(n,H)=(30,3)$}.}
 \label{empty_N30m3}
\end{center}
\end{figure}

\begin{figure}[ht]
\begin{center}
\includegraphics[width=6in]{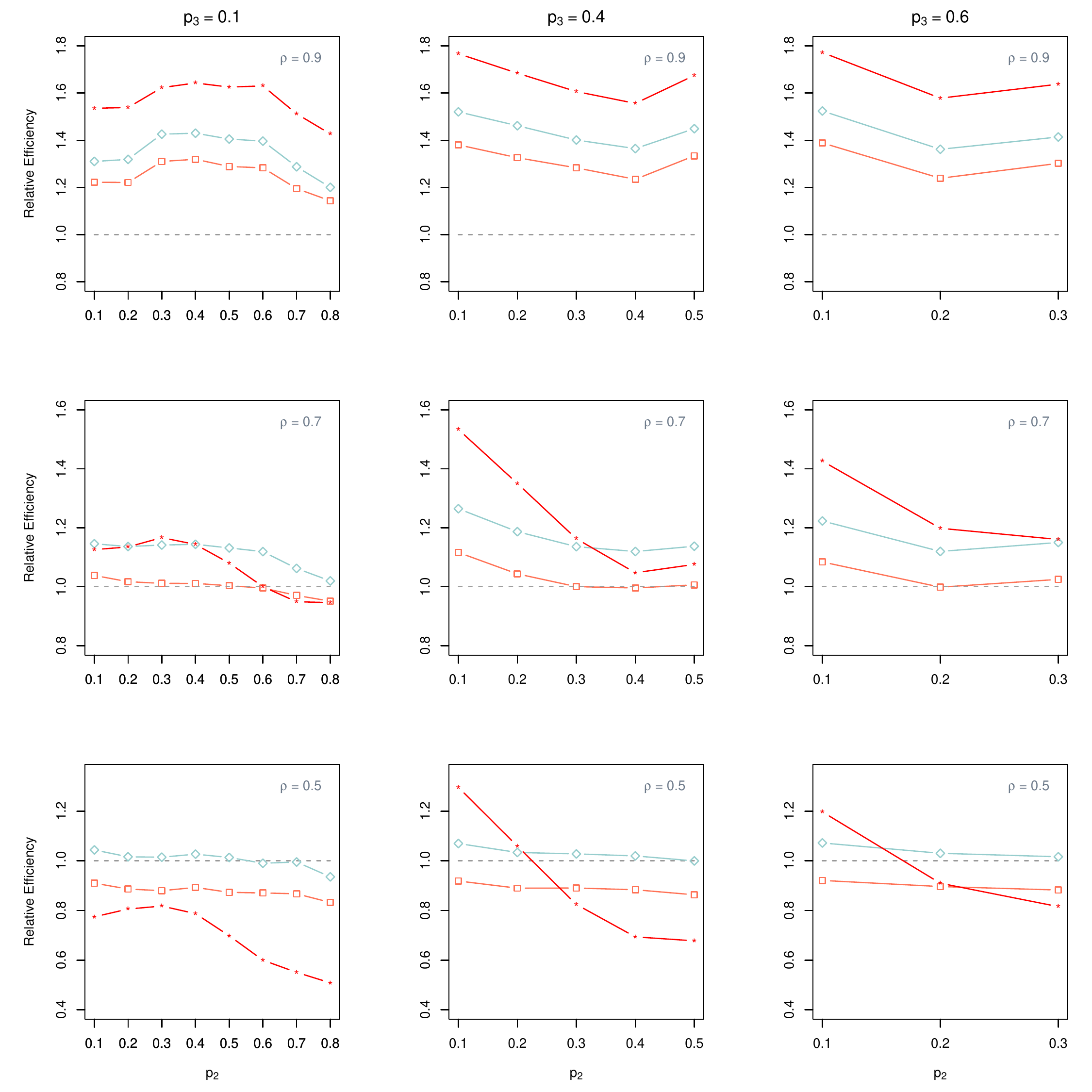}
\caption{\footnotesize{The total REs of $\widehat{\bf p}_{st}$ ({$\color{Melon}\medsquare$}), $\widehat{\bf p}_{iso}$ ({$\color{Aquamarine}\meddiamond$}), $\widehat{\bf p}_{ml}$ ({$\color{Red}\filledstar$}) from JPS without empty strata to their SRS counterpart when $(n,H)=(30,6)$}.}
 \label{no_empty_N30m6}
\end{center}
\end{figure}

\begin{figure}[ht]
\begin{center}
\includegraphics[width=6in]{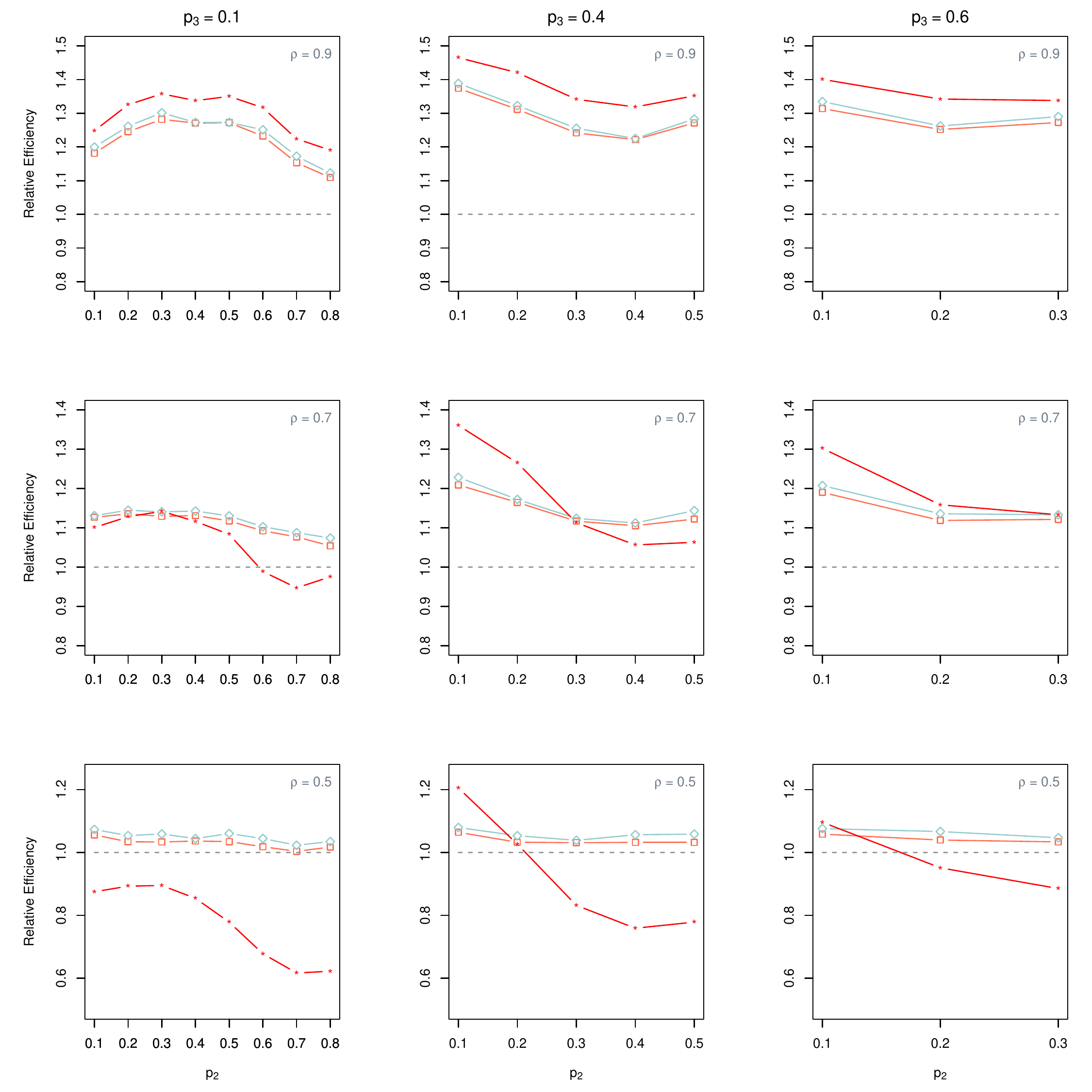}
\caption{\footnotesize{The total REs of $\widehat{\bf p}_{st}$ ({$\color{Melon}\medsquare$}), $\widehat{\bf p}_{iso}$ ({$\color{Aquamarine}\meddiamond$}), $\widehat{\bf p}_{ml}$ ({$\color{Red}\filledstar$}) from JPS without empty strata to their SRS counterpart when $(n,H)=(60,3)$}.}
 \label{no_empty_N60m3}
\end{center}
\end{figure}

\begin{figure}[ht]
\begin{center}
\includegraphics[width=6in]{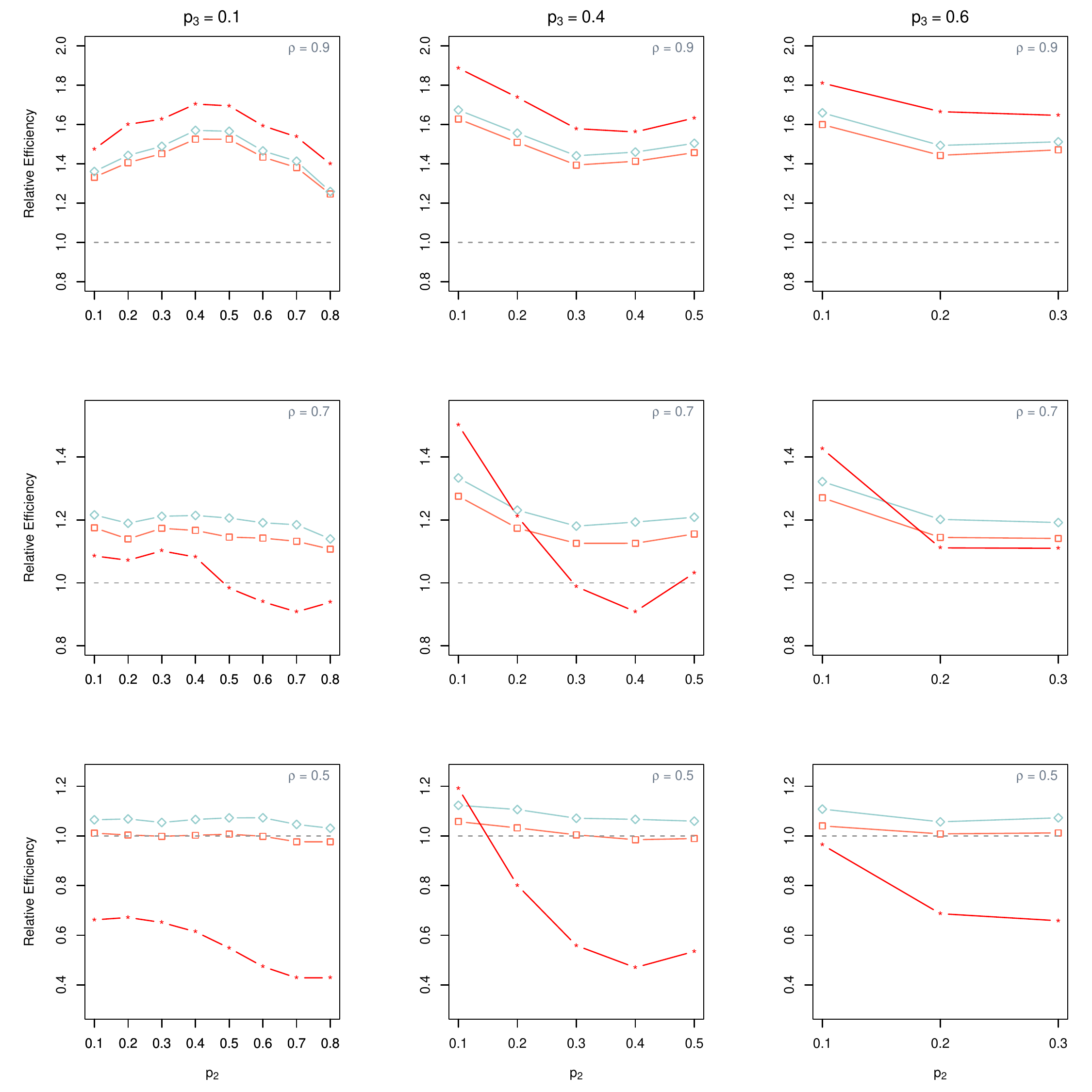}
\caption{\footnotesize{The total REs of $\widehat{\bf p}_{st}$ ({$\color{Melon}\medsquare$}), $\widehat{\bf p}_{iso}$ ({$\color{Aquamarine}\meddiamond$}), $\widehat{\bf p}_{ml}$ ({$\color{Red}\filledstar$}) from JPS without empty strata to their SRS counterpart when $(n,H)=(60,6)$}.}
 \label{no_empty_N60m6}
\end{center}
\end{figure}

\begin{figure}[ht]
\begin{center}
\includegraphics[width=6in]{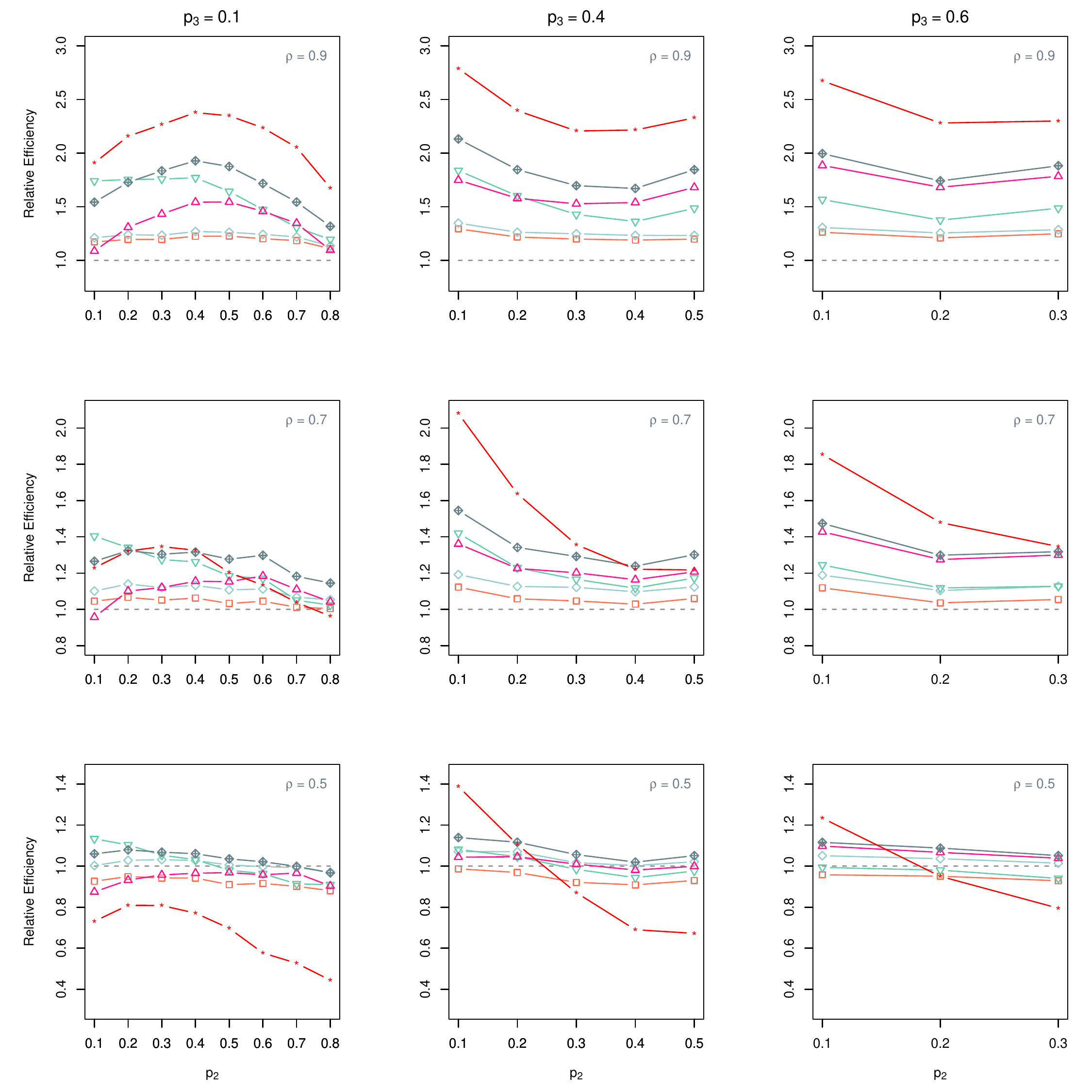}
\caption{\footnotesize{The total REs of $\widehat{\bf p}_{st}$ ({$\color{Melon}\medsquare$}), $\widehat{\bf p}_{iso}$ ({$\color{Aquamarine}\meddiamond$}), $\widehat{\bf p}_{q,{iso}^-}$ ({$\color{JungleGreen}\medtriangledown$}), $\widehat{\bf p}_{q,{iso}^+}$ ({$\color{RubineRed}\medtriangleup$}), $\widehat{\bf p}_{q,{iso}^*}$ ({$\color{RoyalBlue}\diamondplus$}),  $\widehat{\bf p}_{ml}$ ({$\color{Red}\filledstar$}) from JPS with empty strata to their SRS counterpart  when $(n,H)=(30,6)$}.}
 \label{empty_N30m6}
\end{center}
\end{figure}

\begin{figure}[ht]
\begin{center}
\includegraphics[width=6in]{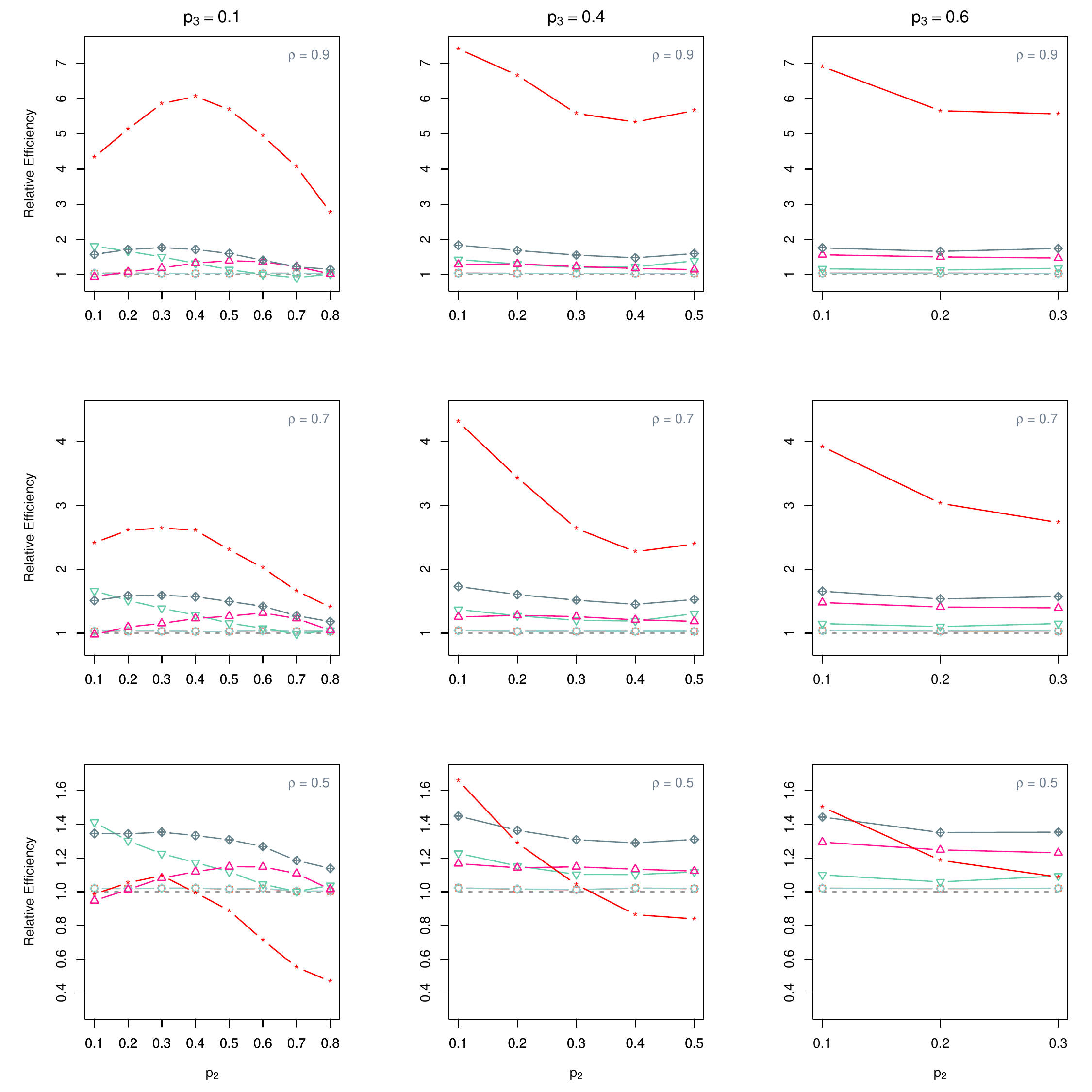}
\caption{\footnotesize{The total REs of $\widehat{\bf p}_{st}$ ({$\color{Melon}\medsquare$}), $\widehat{\bf p}_{iso}$ ({$\color{Aquamarine}\meddiamond$}), $\widehat{\bf p}_{q,{iso}^-}$ ({$\color{JungleGreen}\medtriangledown$}), $\widehat{\bf p}_{q,{iso}^+}$ ({$\color{RubineRed}\medtriangleup$}), $\widehat{\bf p}_{q,{iso}^*}$ ({$\color{RoyalBlue}\diamondplus$}),  $\widehat{\bf p}_{ml}$ ({$\color{Red}\filledstar$}) from JPS with empty strata to their SRS counterpart  when $(n,H)=(60,3)$}.}
 \label{empty_N60m3}
\end{center}
\end{figure}

\begin{figure}[ht]
\begin{center}
\includegraphics[width=6in]{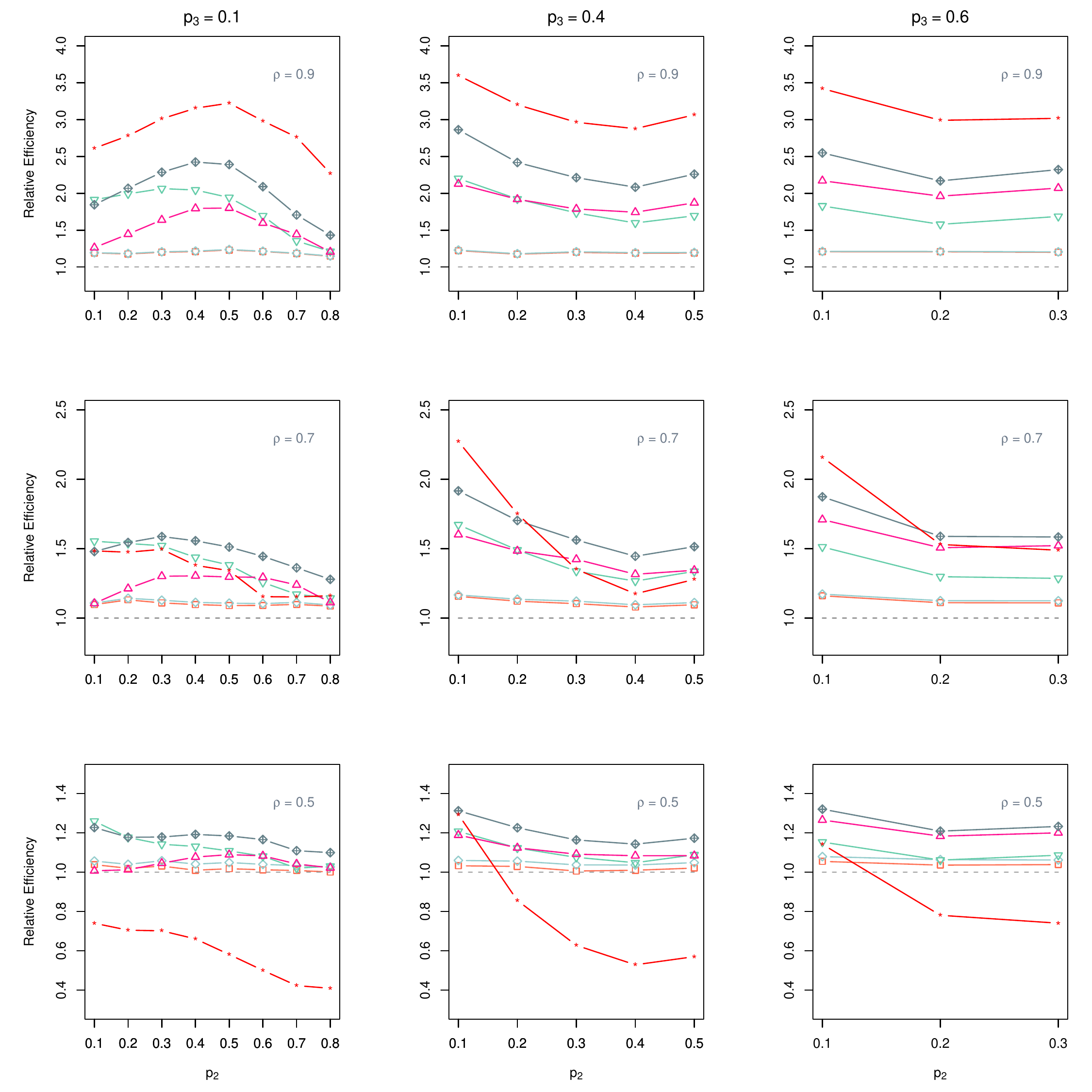}
\caption{\footnotesize{The total REs of $\widehat{\bf p}_{st}$ ({$\color{Melon}\medsquare$}), $\widehat{\bf p}_{iso}$ ({$\color{Aquamarine}\meddiamond$}), $\widehat{\bf p}_{q,{iso}^-}$ ({$\color{JungleGreen}\medtriangledown$}), $\widehat{\bf p}_{q,{iso}^+}$ ({$\color{RubineRed}\medtriangleup$}), $\widehat{\bf p}_{q,{iso}^*}$ ({$\color{RoyalBlue}\diamondplus$}),  $\widehat{\bf p}_{ml}$ ({$\color{Red}\filledstar$}) from JPS with empty strata to their SRS counterpart  when $(n,H)=(60,6)$}.}
 \label{empty_N60m6}
\end{center}
\end{figure}

Figures \ref{no_empty_N30m3} and \ref{empty_N30m3} represent the performance of the estimators using JPS without empty strata and JPS with empty strata, respectively. 
It is observed that the REs of JPS estimators (both ML and non-parametric methods) are almost always greater than one. 
That is, one can more accurately estimate the ordinal population with JPS data than SRS data of the same size.
 The effect of ranking ability $\rho$ is evident  on the performance of the JPS estimators.
When ranking ability is strong, the RE of the JPS estimators significantly grows as the set size $H$ increases from 3 to 6; However, when the ranking ability  is poor, the performance of JPS estimators deteriorates and RE becomes almost one. In other words, when ranking is unreliable, the JPS estimators perform similar to their SRS counterparts in the estimation of the ordinal population (see Figures \ref{no_empty_N30m6} - \ref{empty_N60m6}). 
From Figure \ref{empty_N30m3}, we see the RE of the ML estimators dramatically increase and  they considerably outperform all the non-parametric estimators so that the improvement of non-parametric JPS estimators is  not noticeable at all. 
This dramatic superiority of the MLEs is compatible with the results of \citep{zamanzade2017estimation} (in estimation of binary proportion  from JPS data). This dramatic increase is because of the dependence of MLEs to population configuration. When ranking is high and population parameters are away enough from zero, the MLEs are highly recommended. Otherwise, we observe that non-parametric estimators are more robust (than MLEs) to ranking errors and population configuration. 
We observe the $\widehat{\bf p}_{q,{iso}^*}$ estimate more accurately the population than other nonparametric methods when JPS has empty strata.  This superiority of $\widehat{\bf p}_{q,{iso}^*}$ refers to this fact that it benefits from both $\widehat{\bf p}_{q,{iso}^-}$ and $\widehat{\bf p}_{q,{iso}^+}$ in the estimation. 
 
\begin{figure}[!ht]
\begin{center}
\includegraphics[width=6in]{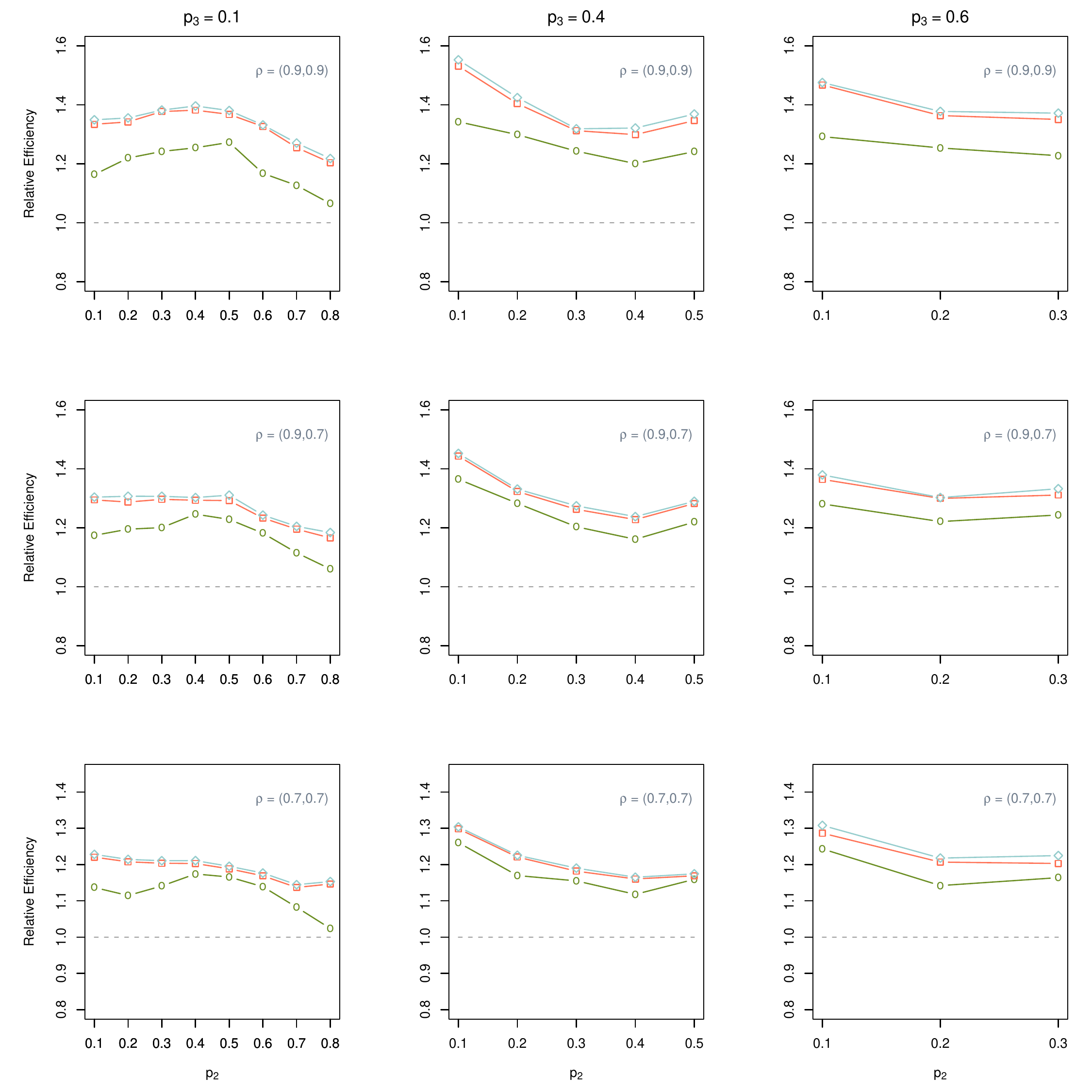}
\caption{\footnotesize{The total REs of $\widehat{{\bf p}}_{reg}$ ({$\color{OliveGreen}\medcircle$}),
 $\widehat{\bf p}_{sm}$ ({$\color{Melon}\medsquare$}), $\widehat{\bf p}_{sm*}$ ({$\color{Aquamarine}\meddiamond$}) from JPS with multiple rankers to their SRS counterpart  when $(n,H)=(30,3)$}.}
 \label{jps_multi_N30m3}
\end{center}
\end{figure}

\begin{figure}[!ht]
\begin{center}
\includegraphics[width=6in]{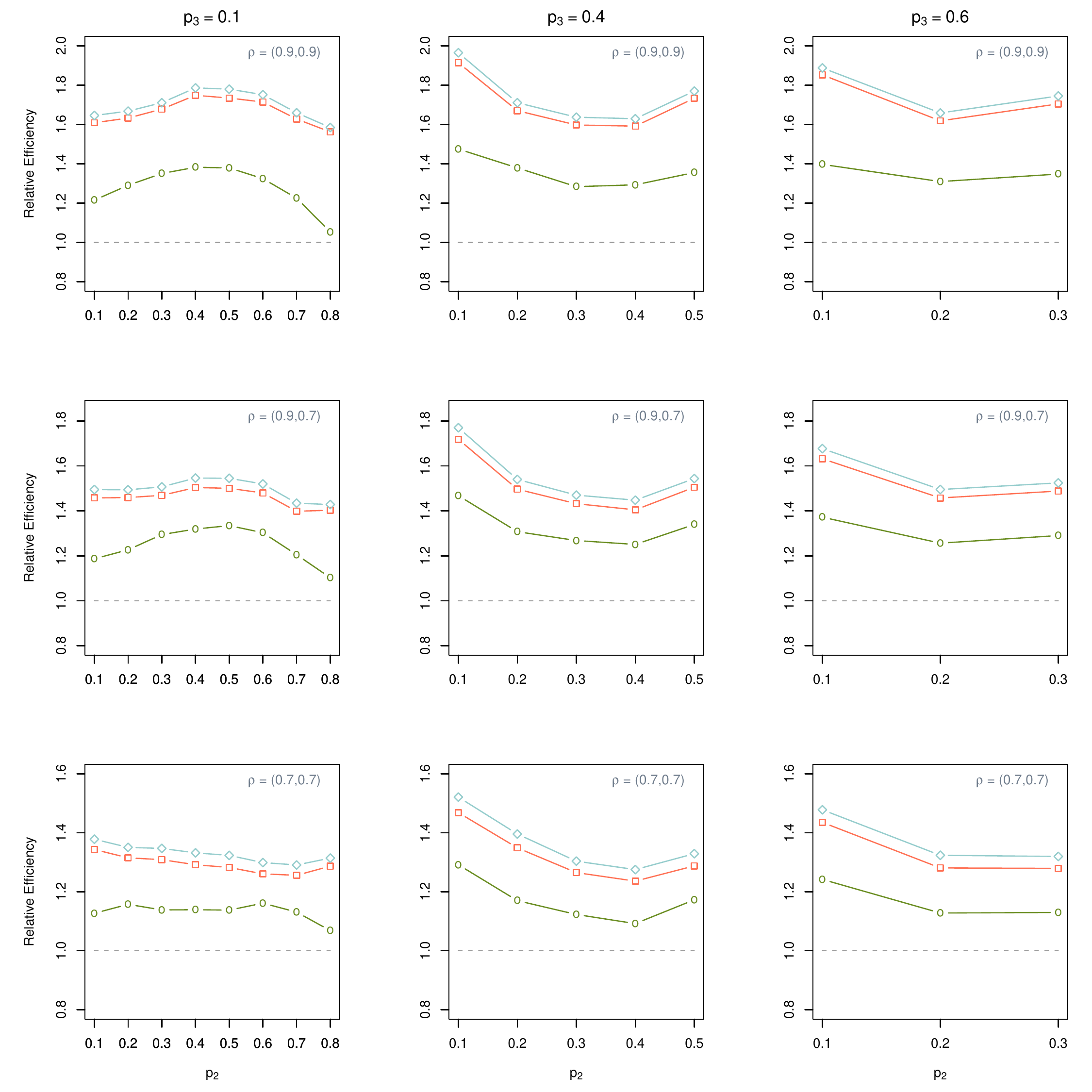}
\caption{\footnotesize{The total REs of $\widehat{{\bf p}}_{reg}$ ({$\color{OliveGreen}\medcircle$}),
 $\widehat{\bf p}_{sm}$ ({$\color{Melon}\medsquare$}), $\widehat{\bf p}_{sm*}$ ({$\color{Aquamarine}\meddiamond$}) from JPS with multiple rankers to their SRS counterpart  when $(n,H)=(30,6)$}.}
 \label{jps_multi_N30m6}
\end{center}
\end{figure}

 Figures \ref{jps_multi_N30m3} and \ref{jps_multi_N30m6} show the results of the estimations from JPS with multiple rankers.  It is seen that all estimators based on multi-ranker JPS outperform their SRS counterparts. 
The standard and the isotonized multi-ranker  estimators perform better than OLR-based counterparts in the estimation of the parameters. 
The shortcoming of the OLR-based method may stem from the fact that the initial JPS sample was used in both training and prediction phases. Consequently, this reduces the ranking ability of the trained OLR model in the comparison sets  in the case of new $H-1$ individuals (who were randomly selected from the population and were not used in the training phase). 
We see the performance of multi-ranker estimators improves as the set size increases and ranking ability is decent.
Unlike OLR-based estimator, the standard and isotonized multi-ranker estimators take full advantage of the high ranking ability and large set size so that their REs  grow considerably in these situations. We also observe the   isotonized multi-ranker estimator almost always outperforms  its standard counterpart.  

\section{Real Data Analysis}\label{sec:real}
In this section, we applied the developed methods for analysis of bone disorder of women aged 50 and older.
Here, the JPS samples are obtained from bone mineral density (BMD) data where measuring the variable interest is expensive and time-consuming. 
Ranking in JPS data is carried out by using some characteristics associated with BMD measurement. Thus, ranking errors arise naturally in JPS data. 

Osteoporosis is a bone metabolic disorder which occurs when the density of the bone tissues deteriorates significantly.  This deterioration  leads to numerous  major health problems including skeletal fragility and  osteoporotic fractures (such as hip). 
For example, more than 50 \% of patients suffering from osteoporotic hip fracture 
are not able to live independently; while around 28\% of these patients will survive less than a year since the complication of the broken bones \citep{neuburger2015impact,bliuc2009mortality,harvey2010osteoporosis}. 
Prevalence of osteoporotic fractures increases as the age increases. Hence, it is critical to monitor and diagnose the  bone disorder status of the population, particularly for the aged population. 

According to the expert panel of World Health Organization, bone mineral density (BMD) is one the most reliable predictors for osteoporosis diagnosis. 
BMD measurements are obtained via dual X-ray absorptiometry (DXA) imaging from various parts of body such as femoral neck. Once the DXA images are acquired,  they must be investigated by medical experts for manual segmentation and quantification of  the final scores. The process of measuring the BMDs is costly and time-consuming. 
BMD measurements are usually reported as T-score (a standardized  score) showing the number standard deviations (SDs) from the BMD norm of the population. The BMD norm is calculated from the BMD measurements of young and healthy individuals between 20 - 30 years old. 
Using the entire information of the population,
the bone disorder status of a patient is diagnosed as osteoporosis when the T-score is more than 2.5 SDs below the BMD norm i.e., $T \le 0.55$. The bone status is determined  as osteopenia  when  $ 0.55 < T \le 0.79$ and the status is considered normal when $T > 0.79$ \citep{kanis2002diagnosis}. Thus, the bone disorder can be represented by an ordinal population with three categories. 

In this study, we work with BMD data from National Health and Nutrition Examination Survey (NHANES III). The survey was  conducted by Centres 
for Disease Control 
and Prevention (CDC) on 33999 Americans between years 1988 - 1994. The Survey consists of two 
examinations of the BMD measurements  from 234 women aged 50 and older. 
 The examinations consist of the BMD measurements from different regions including femur neck (FNBMD), trochanter region (TRBMD) and inter-trochanter region (INBMD).  In this analysis, we treated these 234 individuals as our underlying population.
We also considered the FNBMD from the second examination as our variable of interest $X$ which follows an ordinal distribution with proportions ${\bf p}=(p_1,p_2,p_3)$ with
${p_3}=\P(\text{Normal})=0.26,  {p_2}=\P(\text{Osteopenia})=0.66$ and ${p_1}=\P(\text{Osteoporosis})=0.08$.
To better assess the ranking error, we ranked the patients based on two characteristics $(Z_1,Z_2)$ including  
FNBMD and TRBMD of patients from the first examination with correlations $(\rho_1,\rho_2)= (0.86,0.70)$.

\setlength{\tabcolsep}{5pt}
\begin{table}[h]
      \caption{\footnotesize{The REs of JPS estimators relative to their SRS counterpart when $n =\{30,60\}$ and $H = \{3,6\}$}.}
\vspace{0.3cm} 
\centering 
{\footnotesize
\begin{tabular}{c c c c c c c c c c c c c c c c }
\hline
    \multirow{ 2}{*}{Ranker} &  \multirow{ 2}{*}{$\boldmath{n}$} & \multirow{ 2}{*}{$\boldmath{H}$} &&  
    \multicolumn{6}{c}{ {with at least one empty strata}} &  &
    \multicolumn{5}{c}{{ without empty strata}} 
    \\
    \cline{5-10}
    \cline{12-16}
     & & && $\widehat{\bf p}_{q,st}$ &$\widehat{\bf p}_{q,iso}$ & $\widehat{\bf p}_{q,{iso}^-}$ & $\widehat{\bf p}_{q,{iso}^+}$ 
     & $\widehat{\bf p}_{q,{iso}^*}$ & $\widehat{\bf p}_{q,{ml}}$ &  & $\widehat{\bf p}_{q,st}$ && $\widehat{\bf p}_{q,iso}$ && 
     $\widehat{\bf p}_{q,ml}$\\
    \hline
    \multirow{4}{*}{FNBMD} & \multirow{2}{*}{30} & 3 && 1.06 & 1.06 & 1.02 & 1.36 & 1.36 & \bf{2.41} &  & 1.14 && 1.15 && \bf{1.27} \\

     & & 6 && 1.15 & 1.19 & 1.29 & 1.37 & 1.54 & \bf{1.96} &  & 1.19 && 1.29 & &\bf{1.51} \\
    \cline{2-16}
   &  \multirow{2}{*}{60} & 3 && 1.04 & 1.04 & 0.99 & 1.34 & 1.36 & \bf{4.02} &  & 1.18 && 1.19 && \bf{1.25} \\

     & & 6 && 1.18 & 1.18 & 1.47 & 1.48 & 1.83 & \bf{2.62} &  & 1.36 && 1.40 && \bf{1.54} \\

         \hline
    \multirow{4}{*}{TRBMD} & \multirow{2}{*}{30} & 3 && 1.03 & 1.03 & 1.00 & 1.24 & 1.28 & \bf{1.43} &  & 1.04 && \bf{1.06} && 1.02 \\

     & & 6 && 1.02 & 1.01 & 1.11 & 1.17 & \bf{1.25} & 1.20 & & 1.03 && \bf{1.09} && 1.05  \\
    
    \cline{2-16}
   &  \multirow{2}{*}{60} & 3 && 1.03 & 1.03 & 1.04 & 1.31 & 1.37 & \bf{1.89}  &  & 1.10 && \bf{1.11} && 1.03  \\

     & & 6 && 1.08 & 1.10 & 1.22 & 1.32 & \bf{1.46} & 1.22 &  & 1.11 && \bf{1.16} && 1.01\\
   \hline
    \end{tabular}
    }   
         \label{tab_single}
\end{table}

We selected JPS data of size $n=\{30,60\}$ with set size $H=\{3,6\}$ via sampling with replacement from the bone disorder population. For each combination $(n,H)$, we replicated 5000 times the JPS data collection
and computed the total REs \eqref{tre} of the developed  JPS estimators relative to their  SRS counterpart of the same
 size. Similar to simulations studies, we use the initial SRS samples to train the OLR models.
The results of the empirical study are illustrated in Tables \eqref{tab_single} and \eqref{tab_mult}. It is apparent 
that the REs of all the JPS estimators are almost always greater than 1. Thus, JPS methods outperform their commonly
 used SRS counterpart in the estimation of bone disorder population.
From Table \eqref{tab_single}, when ranking ability is decent in JPS (with and without empty strata), the REs 
of the JPS estimators improve. This improvement grows when set size increases from 3 to 6. In addition, ML estimators
 always outperform the non-parametric estimators. This superiority is compatible with what we observed in simulation
  studies that RE of the ML estimators dramatically increase when ranking ability is high and population proportions 
  are away from zero.  Comparing between the non-parametric estimators, the $\widehat{\bf p}_{{iso}^*}$ appear more 
  promising in the estimation of bone disorder population. From Table \eqref{tab_mult},  the standard and isotonized multi-rankers
  perform better than their OLR-based counterpart. Finally, the isotonized multi-ranker estimator is recommended to practitioners to estimate the parameters of the bone disorder population. 

\setlength{\tabcolsep}{5pt}
\begin{table}[h]
      \caption{\footnotesize{The REs of JPS estimators using multiple rankers relative to their SRS counterpart when $n =\{30,60\}$ and $H = \{3,6\}$}.}
\vspace{0.3cm} 
\centering 
{\footnotesize
\begin{tabular}{c c c c c c  c  c c  c  c }
\hline
     {Rankers} & & $n$ & & $H$ & &$\widehat{\bf p}_{sm}$ & & $\widehat{\bf p}_{{\ah{sm*}}}$ & & $\widehat{\bf p}_{{reg}}$\\\hline
    \multirow{4}{*}{FNBMD $\boldsymbol{\&}$ TRBMD} & & \multirow{2}{*}{30} & & 3 & & 1.18 & & ${\bf 1.20}$ & & 1.11 \\
    & & & & 6 && 1.36 & & ${\bf 1.40}$ & & 1.15 \\
    \cline{3-11}
   & & \multirow{2}{*}{60} & & 3 && 1.24 & & ${\bf 1.26}$ & & 1.20  \\
     & & & & 6 && 1.46 & & ${\bf 1.48}$ & & 1.31 \\
   \hline
    \end{tabular}
    }      
         \label{tab_mult}
\end{table}

\section{Summary and Concluding Remarks}\label{sec:sum}
In many applications, measuring the variable of interest is difficult; 
however, there may be easy to measure characteristics about the sampling units. They can be effectively exploited 
for ranking and data collection process.  Ranked set sampling and judgment post-stratification sampling 
are two cost-effective sampling schemes  which use these characteristics as ranking information and enable us to obtain more representative sampling from the variable of interest. 

Although \cite{chen2008ranked} and \cite{alvandi2021estimation} have used RSS data for estimation of ordinal proportions, RSS shares various challenges in the data collection. For example, RSS data structure differs from the population structure and ranking information can not be separated from the RSS observations. 
Unlike RSS, JPS sampling results in random sample from the population. If ranking is unreliable, we can easily ignore the ranks. Thus, we can treat the JPS data as SRS data and apply the conventional statistical methods  to JPS observations. 

In this paper, to our best knowledge, we study for the first time the properties  of JPS samples for categorical ordinal variables.  
 We developed various estimators from JPS data. These estimators include three estimators based on JPS without empty strata, six estimators from JPS with empty strata and three JPS estimators from multiple rankers.  Through simulation studies, we investigated the effect of ranking error, set size, population configuration on performance of the estimators.
   The developed estimators were then applied to an empirical example for analysis of bone disorder status of the aged population.  
  From numerical investigation, when practitioners have access to a decent ranker, ML estimators are recommended for the estimation of the ordinal population; Otherwise, we recommend isotonized non-parametric estimators like $\widehat{\bf p}_{{iso}^*}$, which are more robust to ranking ability and population configuration.

\section*{Acknoledgments}
Armin Hatefi acknowledges the research support of the Natural Sciences and Engineering Research Council of Canada (NSERC).

\bibliographystyle{plainnat}
\bibliography{MultJPS-bib}

\newpage
\section*{Appendix}

\subsection*{Proof of Equation \eqref{ll}:}
The following shows 
how the likelihood function given ranks in equation \eqref{ll} is constructed. Note that

\begin{itemize}
\item[a)] $X_1,\ldots,X_n$ are iid random variables from multinomial distribution with $Q$ ordinal categories; that is $ X_i \overset{iid}{\sim} Multi(1,p_1,\ldots,p_Q)$.
\item[b)] Let $Y_{(h)}= (X_i | R_i=h)$, for $i=1,\ldots,n$. In perfect JPS, $Y_{(h)}$ is the $h$-th order statistic in the $i$-th comparison set of size $H$.
\item[c)] $Y_{(1)},\ldots,Y_{(n)}$ are independent order statistics since they are come from independent comparison sets. 
\item[d)] $Y_{(h)} \sim Multi(1,p_{(h)1},\ldots,p_{(h)Q})$ where
\end{itemize}
\begin{align*}
p_{(h)q}  =& P(Y_{(h)} = q) = P (Y_{(r)} \le q) - P (Y_{(r)} \le q-1) \\
& P( \text{at least $h$ units in the set of size $H$ come from categories $\{1,\ldots,q\}$})  \\
&- P( \text{at least $h$ units in the set of size $H$ come from categories $\{1,\ldots,q-1\}$}) \\
=& \sum_{t=h}^{H}  {H \choose t} c_q^t (1-c_q)^{H-t} - \sum_{t=h}^{H}  {H \choose t} c_{q-1}^t (1-c_{q-1})^{H-t} \\
=& B_{c_q}(h,H-h+1) - B_{c_{q-1}}(h,H-h+1) \\
=& \frac{1}{B(h,H-h+1)} \left( \int_{0}^{c_q} u^{h-1} (1-u)^{H-h} du - \int_{0}^{c_{q-1}} u^{h-1} (1-u)^{H-h} du\right) \\
=& \frac{1}{B(h,H-h+1)}  \int_{c_{q-1}}^{c_q} u^{h-1} (1-u)^{H-h} du.
\end{align*}
From (c) and (d), the likelihood function given the ranks is given by
\[
L(c_1,\ldots,c_{Q-1}) = \prod_{i=1}^{n}  \left[p_{(r_i)1}\right]^{\I(X_i=1)}    \left[p_{(r_i)2}\right]^{\I(X_i=2)} \cdots \left[1-\sum_{q=1}^{Q-1}p_{(r_i)q}\right]^{\I(X_i=Q)}.
\]
Consequently,  the log-likelihood function (i.e., equation \eqref{ll} in the manuscript) is constructed as
\[
\ell(c_1,\ldots,c_{Q-1})=\sum_{i=1}^{n} 
\left\{ 
 \I(X_i=1) \log(p_{(r_i)1}) + \ldots +\I(X_i=Q) \log\left(1- \sum_{q=1}^{Q-1}p_{(r_i)q}\right)  
\right\}.
\]

  \begin{figure}[ht]
\begin{center}
\includegraphics[width=6in]{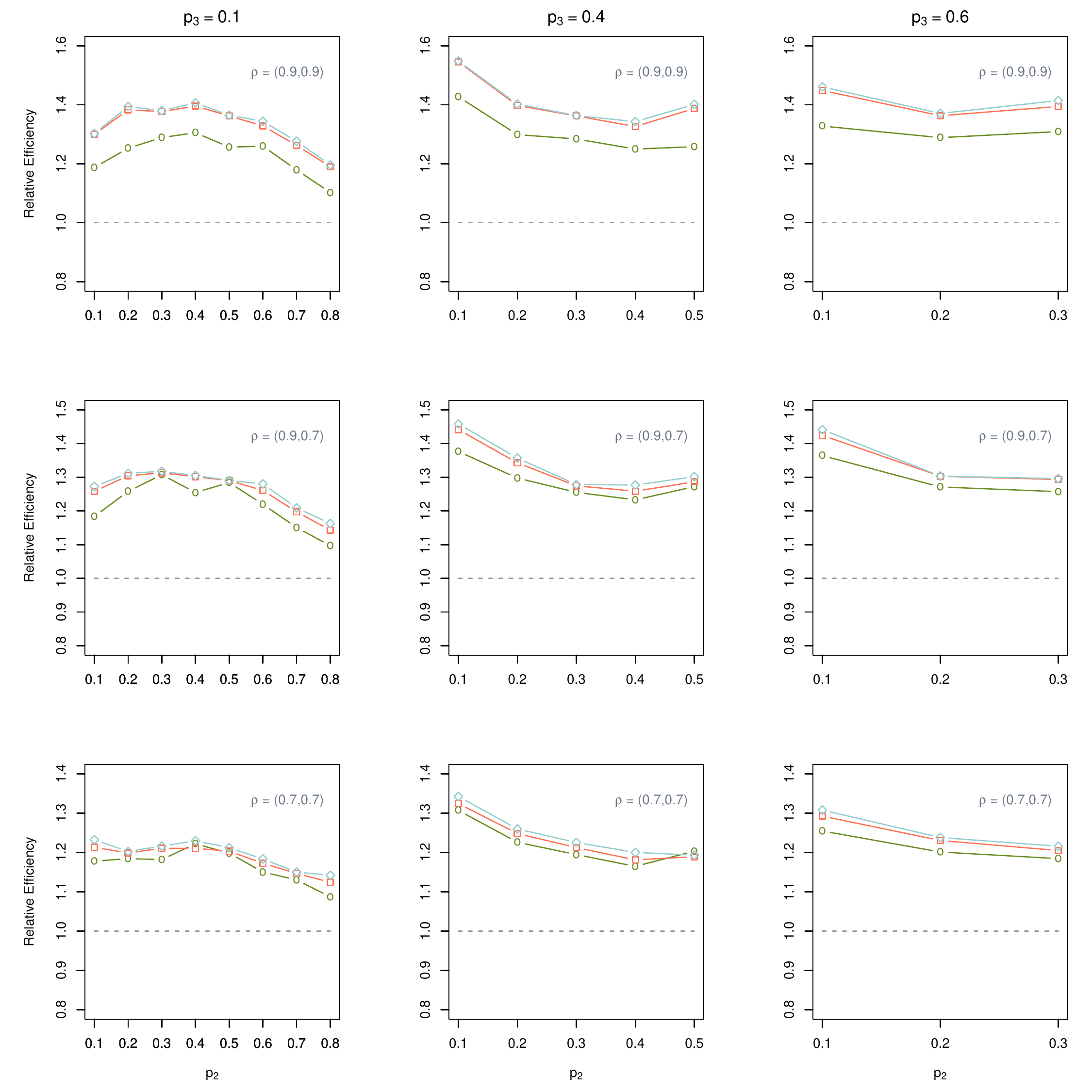}
\caption{\footnotesize{The total REs of $\widehat{{\bf p}}_{reg}$ ({$\color{OliveGreen}\medcircle$}),
 $\widehat{\bf p}_{sm}$ ({$\color{Melon}\medsquare$}), $\widehat{\bf p}_{sm*}$ ({$\color{Aquamarine}\meddiamond$}) from JPS with multiple rankers to their SRS counterpart  when $(n,H)=(60,3)$}.}
 \label{jps_multi_N60m3}
\end{center}
\end{figure}

  \begin{figure}[h]
\begin{center}
\includegraphics[width=6in]{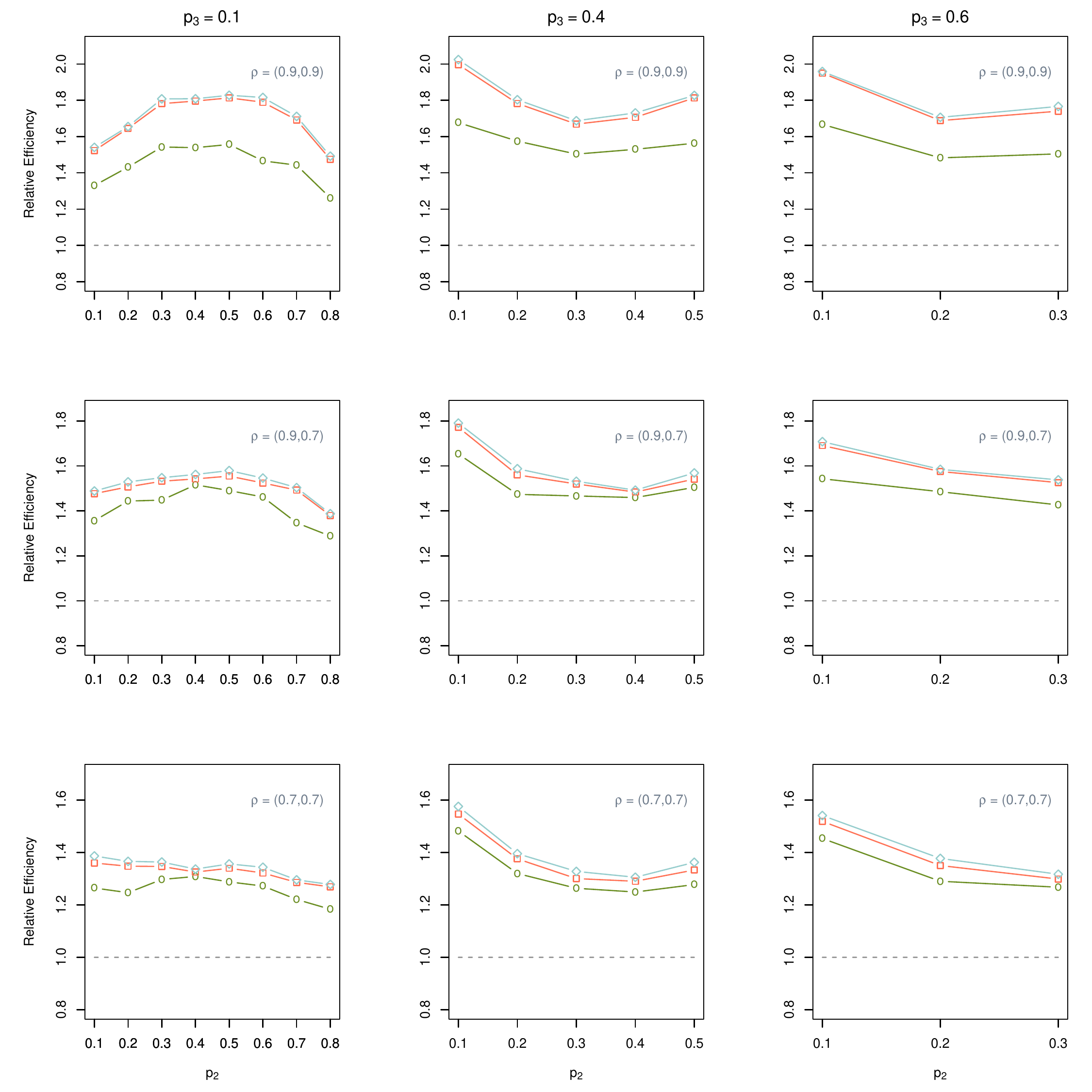}
\caption{\footnotesize{The total REs of $\widehat{{\bf p}}_{reg}$ ({$\color{OliveGreen}\medcircle$}),
 $\widehat{\bf p}_{sm}$ ({$\color{Melon}\medsquare$}), $\widehat{\bf p}_{sm*}$ ({$\color{Aquamarine}\meddiamond$}) from JPS with multiple rankers to their SRS counterpart  when $(n,H)=(60,6)$}.}
 \label{jps_multi_N60m6}
\end{center}
\end{figure}

%
\end{document}